\documentclass[aps,prl,twocolumn,superscriptaddress,floats]{revtex4}
\usepackage{txfonts}
\usepackage{amssymb}
\usepackage{graphicx}

\begin{document}

\title{Electronic identification of the actual parental phase of  K$_x$Fe$_{2-y}$Se$_2$ superconductor and its intrinsic mesoscopic phase separation}

\author{F. Chen}\author{M. Xu}\author{Q. Q. Ge}\affiliation{State Key Laboratory of Surface Physics, Department of Physics,  and Advanced Materials Laboratory, Fudan
University, Shanghai 200433, People's Republic of China}

\author{Y. Zhang}
\email{yanzhangfd@fudan.edu.cn}
\affiliation{State Key Laboratory of Surface Physics, Department of Physics, and Advanced Materials Laboratory, Fudan University, Shanghai 200433, People's Republic of China}

\author{Z. R. Ye}\author{L. X. Yang}\author{Juan Jiang}\author{B. P. Xie}
\affiliation{State Key Laboratory of Surface Physics, Department of Physics,  and Advanced Materials Laboratory, Fudan
University, Shanghai 200433, People's Republic of China}

\author{R. C. Che}\affiliation{Department of Materials Science, and Advanced Materials Laboratory, Fudan University, Shanghai 200433, People's Republic of China}

\author{M. Zhang}\author{A. F. Wang}\author{X. H. Chen}
\affiliation{Hefei National Laboratory for Physical Sciences at Microscale and Department of Physics, University of Science and Technology of China, Hefei, Anhui 230026, China}

\author{D. W. Shen}\author{X. M. Xie}\author{M. H. Jiang}
\affiliation{State Key Laboratory of Functional Materials for Informatics, Shanghai Institute of Microsystem and Information Technology, Chinese Academy of Sciences, Shanghai 20005}

\author{J. P. Hu}
\affiliation{Department of Physics, Purdue University, West Lafayette, Indiana 47907, USA}

\author{D. L. Feng}\email{dlfeng@fudan.edu.cn}
\affiliation{State Key Laboratory of Surface Physics, Department of Physics, and Advanced Materials Laboratory, Fudan University, Shanghai 200433, People's Republic of China}

\begin{abstract}
\end{abstract}

\maketitle

\textbf{While the parent compounds of the cuprate high temperature
superconductors (high-$T_c$'s) are Mott insulators \cite{Cuprate}, the
iron-pnictide high-$T_c$'s are in the vicinity of a metallic spin
density wave (SDW) state \cite{Hosono,ChenXH}, which
highlights the difference between these two families.
However, insulating parent compounds were identified for the newly discovered K$_x$Fe$_{2-y}$Se$_2$ \cite{ChenPhase,FangMH1}. This raises an intriguing question as to whether the iron-based high-$T_c$'s could be viewed as doped Mott insulators
like the cuprates. Here we report  angle-resolved photoemission
spectroscopy  (ARPES) evidence of two insulating and one
semiconducting phases of K$_x$Fe$_{2-y}$Se$_2$, and the mesoscopic
phase separation between the superconducting/semiconducting phase
and the insulating phases. The insulating phases are characterized by the depletion of electronic states over a 0.5~eV window below
the chemical potential, giving a compelling evidence for the presence of Mott-like physics. The charging effects and the absence of band folding in the superconducting/semiconducting phase further prove that the static magnetic and vacancy orders are not related to the superconductivity. Instead, the electronic structure of the superconducting phase is much closer to the semiconducting phase, indicating the superconductivity is likely developed by doping the semiconducting phase rather than the insulating phases.}

\begin{figure}[b]
\includegraphics[width=8.7cm]{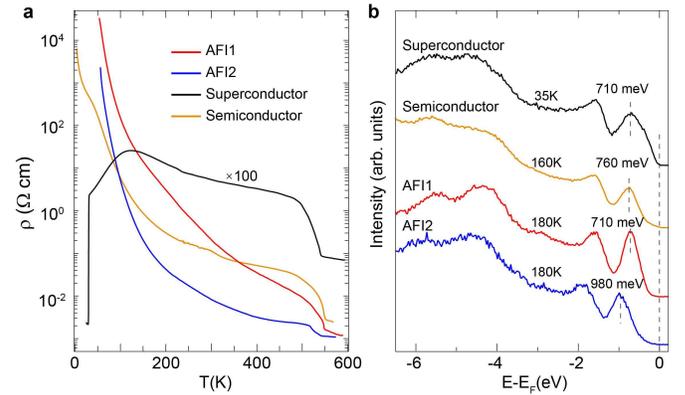}
\caption{\textbf{Resistivities and valence bands of various
K$_x$Fe$_{2-y}$Se$_2$ compounds.} \textbf{a}, In-plane resistivities as a function of temperature for the AFI1 and AFI2 insulating compounds, the semiconducting compound, and the
superconducting compound with the superconducting transition temperature ($T_c$)~$=$31~K. The resistivity of the superconducting sample is multiplied by 100. \textbf{b}, Valence band photoemission spectra at the $\Gamma$ point for the
insulators, semiconductor, and superconductor. The dashed lines are
guides to eyes for the peak positions of the feature near
-1~eV. If not specified otherwise, the data in this figure
and hereafter were taken with
21.2~eV photons from an in-house helium discharge lamp. The data for superconductor, semiconductor, and insulators were measured at
35, 160, and 180~K, respectively.} \label{vb}
\end{figure}

\begin{figure*}[t]
\includegraphics[width=16cm]{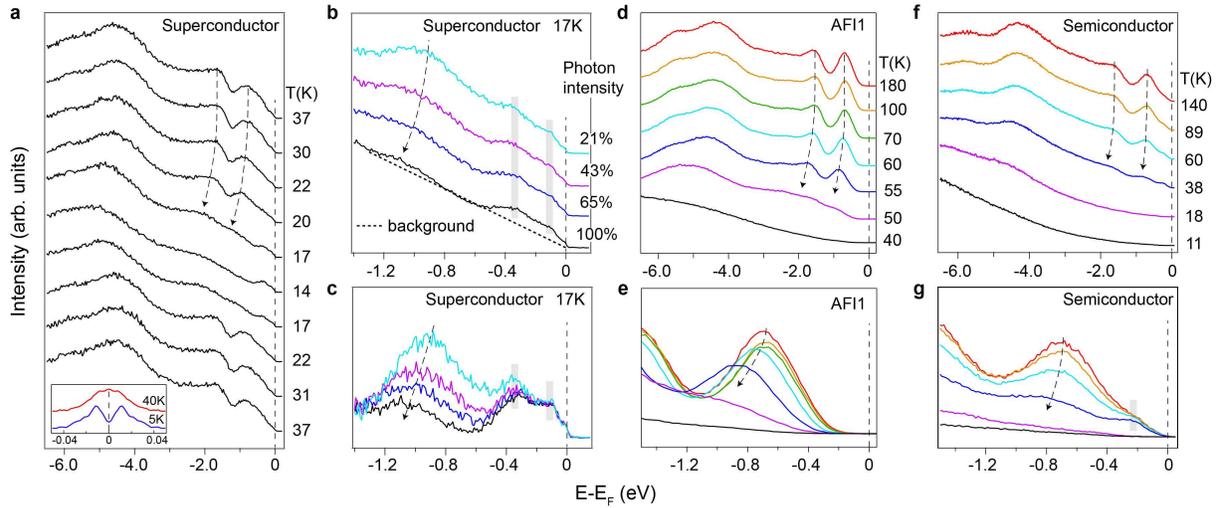}
\caption{\textbf{Photoemission charging effects in K$_x$Fe$_{2-y}$Se$_2$.} \textbf{a}, Temperature
dependence of the photoemission spectrum at the $\Gamma$ point of
superconductor. The inset is the symmetrized EDC's across $T_c$ illustrating a superconducting
gap of about 10~meV at the electron Fermi surface around the zone
corner. \textbf{b}, Energy distribution curves
(EDC's) at the $\Gamma$ point of superconductor as a function of
the photon intensity. Data were taken at 17~K. \textbf{c}, The
EDC's after deducting the background (dashed line in \textbf{b}) to
highlight the peak position shift due to charging effect.  \textbf{d}, Temperature dependence of the EDC's at the
$\Gamma$ point for the AFI1 sample. \textbf{e}, The low energy
portion of data in \textbf{d}. \textbf{f} and \textbf{g} are similar to  \textbf{d} and \textbf{e} respectively, but for the semiconducting sample. The arrows on the dashed lines indicate the band shift with enhanced charging effect.}
\label{cha}
\end{figure*}

The recent discovery of a new family of iron-based superconductor
A$_x$Fe$_{2-y}$Se$_2$ (A= K, Cs, Rb, ...) has raised a lot of
interest as it is rather unique compared with all the other series
of iron-based superconductor \cite{ChenXL, ChenXH2}. ARPES experiment has found the absence of the hole Fermi pockets
near the zone center, thus ruling out the so called s$^{\pm}$
pairing symmetry in this system \cite{ZhangNM}. Moreover, the
superconducting phase is in adjacent with an insulating and
magnetically ordered state instead of a metallic SDW state \cite{ChenPhase,FangMH1}.
Particularly, there are superlattice modulations due to iron vacancy order at high
temperature followed by an antiferromagnetic order. Neutron scattering
studies found that the ordered moment could be as large as
3.3~$\mu_B$  per iron cation \cite{Bao1}, which is the
largest among all the discovered parent compounds of iron-based
superconductors so far. Further thermal power and transmission electron
microscope (TEM) measurements were able to distinguish two
insulating phases in K$_x$Fe$_{2-y}$Se$_2$: an ``AFI1" phase
characterized by a positive thermal power and a superlattice
modulation wave-vector (1/5, 3/5, 0) in the reciprocal space, and an
``AFI2" phase characterized by a negative thermal power and a
superlattice modulation wave-vector (1/4, 3/4, 0)
\cite{ChenPhase, TEM1}. Theoretically, some suggest that the iron vacancy
order would cause band narrowing and drive the system into a Mott
insulator \cite{SiMott}, and thus leading to an  intriguing
possibility that  there could be a united mechanism for both the
copper-based and iron-based high-$T_c$'s. However, others
argue that it could be a simple band insulator \cite{Mazin} since
density-functional theory calculations could reproduce the
experimental antiferromagnetic ordered pattern given the iron vacancy order \cite{Lu15, Lu16}.


To reveal the nature of the parent compounds of this new family of
iron-based superconductor, we have studied a series of
K$_x$Fe$_{2-y}$Se$_2$ with ARPES (details are in the Methods
section).  As illustrated in Fig.~\ref{vb}a, the resistivity
behaviors of these K$_x$Fe$_{2-y}$Se$_2$ samples are rather
versatile, including the insulating and superconducting ones
\cite{ChenPhase}, and a semiconducting behavior that is much less
insulating below 100~K as reported before \cite{WangNLKFe2Se2}.
Interestingly, the resistivity of the superconducting compound exhibits an
insulating behavior above 120~K as well. Moreover, the resistivity anomaly
near 550~K is present in all the samples, indicating the formation of vacancy order \cite{ChenPhase, coex3}. Figure~\ref{vb}b
shows the valence band photoemission spectra near the zone center
($\Gamma$) over a large energy window for all four types different compounds. The main spectral features are
comparable in all cases, however, difference does show up quantitatively. For example, the feature near -1~eV differs by 270~meV for AFI1 and AFI2, indicating these two insulating phases are indeed electronically different. The peak of the semiconducting sample is situated in between AFI1 and AFI2, while the peak position of the
superconducting sample is the same as that of AFI1, and it clearly
contains a shoulder structure near the Fermi energy ($E_F$).


\begin{figure}[b]
\includegraphics[width=8.7cm]{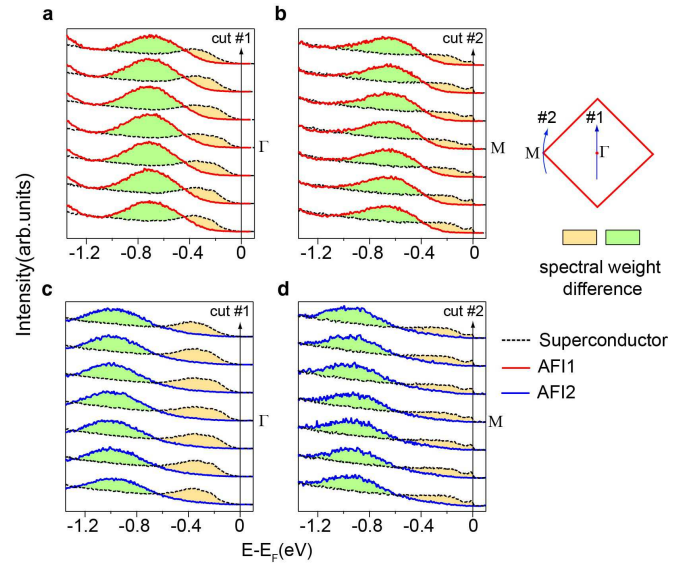}
\caption{\textbf{Spectral weight difference between
the insulators and the superconductor.} \textbf{a}, \textbf{b},
compare EDC's of AFI1 insulator and the superconductor along
cuts \#1 and \#2 across the  projected two dimensional  Brillouin
zone center and corner respectively. \textbf{c}, \textbf{d}, make
the same comparison between AFI2 and the superconductor. Data of insulator and superconductor were taken at 180 and 11~K respectively. } \label{ins}
\end{figure}

\begin{figure*}[t]
\includegraphics[width=16cm]{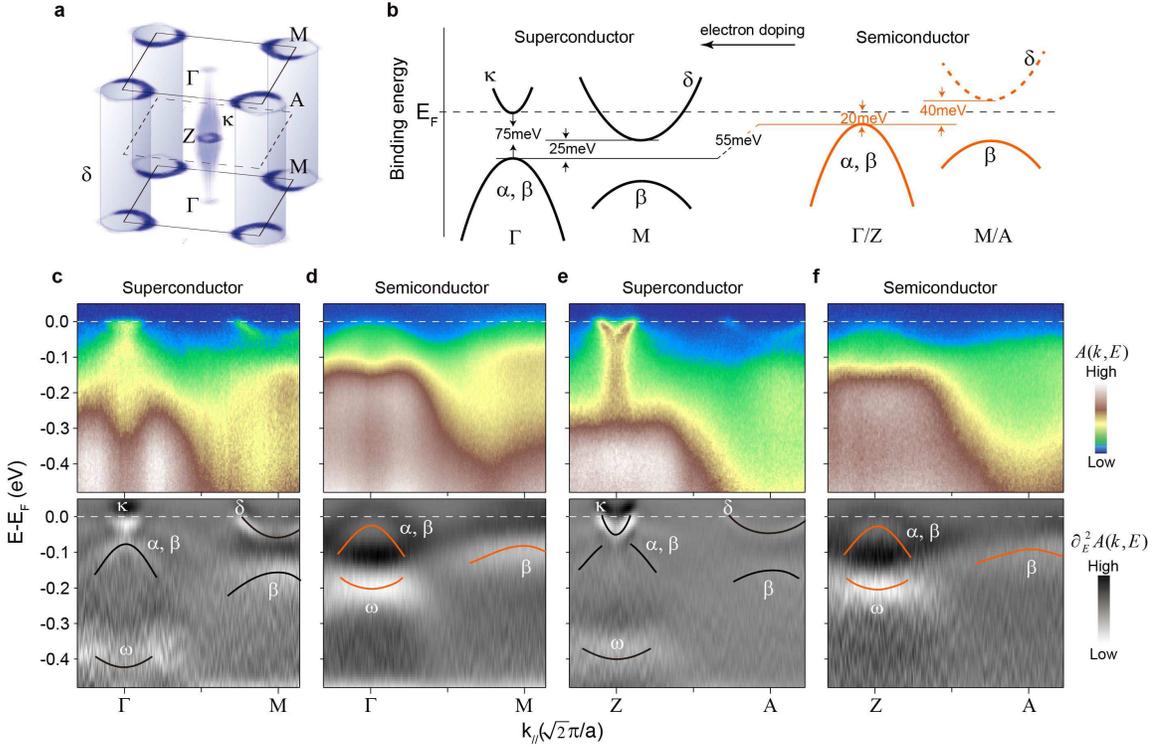}
\caption{\textbf{ Low energy electronic structures of the superconducting and
semiconducting  K$_x$Fe$_{2-y}$Se$_2$.}  \textbf{a}, The Fermi surface of the superconducting phase in the three-dimensional  Brillouin zone. \textbf{b}, The sketch of the band structure evolution from the semiconductor to the superconductor. \textbf{c}, The photoemission intensities (upper panel) and its second derivative with
respect to energy (lower panel) along the $\Gamma$ - $M$ direction for the superconductor taken at 35~K. \textbf{d} is the same as \textbf{c} but for the semiconductor taken at 100~K. Data were taken at SSRL with 21~eV photons. \textbf{e} and \textbf{f} are same as \textbf{c}, \textbf{d}, except the data were taken with 31~eV photons to measure the $Z$-$A$ direction.} \label{smsc}
\end{figure*}

\begin{figure}[t!]
\includegraphics[width=8.7cm]{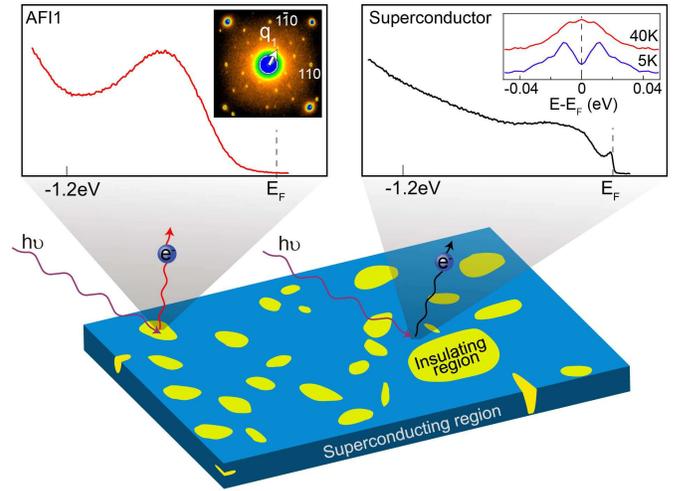}
\caption{\textbf{Cartoon for mesoscopic phase
separation in K$_x$Fe$_{2-y}$Se$_2$.}  Different regions exhibit
different photoemission spectroscopic signature. Diffraction
pattern of the $\sqrt{5} \times \sqrt{5}$ order was
observed with TEM in both the superconductors and semiconductors
indicating a mixing of superconducting or semiconducting phase with
the AFI1 phase. The arrow in diffraction pattern indicates the superlattice modulation wave vector $q_1$~=~(1/5,3/5,0). The TEM data was collected at room temperature. The upper right inset is the symmetrized EDC's of the superconductor across $T_c$, illustrating a superconducting gap.} \label{sep}
\end{figure}

The rather similar valence bands in Fig.~\ref{vb}b raise a question that whether these features are related to the intrinsic property of electronic structures of the drastically different phases of K$_x$Fe$_{2-y}$Se$_2$. With detailed temperature dependence in Fig.~\ref{cha}a, the two features at -1.6 and -0.7~eV in the superconducting sample fade away at 17~K and disappear at 14~K. Such a behavior is reversed with further increasing temperature. Moreover, if we reduce the intensity of the 21.2~eV
photons to reduce the photocurrent, these valence band features
recover (Fig.~\ref{cha}b). As shown in the background-deducted spectra
in Fig.~\ref{cha}c, the valence band moves toward $E_F$ upon
decreasing the incident photon intensity. The reaction of the valence band peaks at -1.6 and
-0.7~eV to the photon intensity and temperature are the typical
charging behaviors of an insulator. The resistivity quickly
increases at lower temperatures. When the photoelectrons leave an
insulating region, the electric charge and potential will build up, which
would smear and shift the photoemission spectrum to higher binding energies (Fig.~\ref{cha}b). However, as the number of photoelectrons is decreased by reducing the photon intensity, the charging behavior could be suppressed (Fig.~\ref{cha}c). Therefore, the high energy charging feature do not belong to the metallic region of the sample, but from the insulating regions in the material, which is the
AFI1 type of insulating region judging from their valence band
positions in Fig.~\ref{vb}b and TEM evidence presented later in
Fig.~\ref{sep}. On the other hand, the low energy features in the first 0.5~eV below $E_F$ show charging-free behavior in Fig.~\ref{cha}c, and are absent in the AFI1 sample (Fig.~\ref{cha}e) indicating that they belong to metallic regions which are phase-separated from the insulating regions. In fact, the inset of Fig.~\ref{cha}a shows that these metallic states open a superconducting gap near $E_F$ below $T_c$ \cite{ZhangNM}. Similar phase separation in the semiconducting sample is illustrated in Figs.~\ref{cha}f and \ref{cha}g. We note that the
insulating feature (represented by features below -0.5~eV) already shows charging behavior at high
temperature, while the semiconducting features (represented by features in the first 0.5~eV below $E_F$) start to show some
charging behavior below 38~K (Figs.~\ref{cha}d and \ref{cha}e) as expected from its increased resistivity. Furthermore, in the AFI1 and semiconducting samples, we found that charging occurs below 70~K (Figs.~\ref{cha}d and \ref{cha}f), while in the superconducting sample, it happens at a much lower temperature of 17~K. Therefore, such a phase separation must happen in a mesoscopic scale.

The prominent contribution of the insulating phase to the valence band suggests that the insulating phase occupies
a rather large fraction of the semiconducting/superconducting samples. Figure~\ref{ins} compares the electronic structures of the insulating compounds with that of the superconductor measured at 11~K, so that
the contribution from the insulating regions to the spectra is
minimized by their own charging. The low energy spectral weight in
the superconducting case seems to be transferred into the
feature around -0.7~eV in Figs.~\ref{ins}a and \ref{ins}b for AFI1, and -1~eV in
Figs.~\ref{ins}c and \ref{ins}d for AFI2. Such a strong spectral weight transfer
resembles the opening of the Mott-Hubbard gap when electron-electron
Coulomb interactions are turned on \cite{Cuprate, SiMott}. Such a Mott-like behavior in
the insulating phases have been recently predicted in LSDA+U (local
spin density approximation plus Coulomb interactions) calculations
for both AFI1 and AFI2  \cite{Dai16,Lu15,Lu16}. It was shown that the vacancy order alone
would not open a large energy gap in the band structure\cite{Lu15,Lu16}. By including the magnetic
order, the band is renormalized greatly. A gap would open for AFI1
even without including U in the calculation as the magnetic order is
extremely strong there, but including U would further increase this gap \cite{Lu16, Dai16}; while U is needed for the AFI2 case to open a gap \cite{Lu15}.  Our observation of the gap in both cases, particularly in the AFI2 case
suggests Coulomb interaction is an important factor of the physics here.
The essentially non-dispersive feature of the insulators in Fig.~\ref{ins} further agrees with the calculations.


In contrast to the huge differences between the electronic structure of the insulating and superconducting phases, the low energy features observed in the semiconducting and superconducting samples show similar charging-free behavior, suggesting the intimate relationship between these semiconducting and superconducting phases. Figure~\ref{smsc} compares their low energy electronic structures along two high symmetry cuts in the three-dimensional Brillouin zone. As illustrated in
Fig.~\ref{smsc}a for the superconductor, there are a $\kappa$ Fermi pocket
around Z, and four $\delta$ Fermi cylinders around the zone corner edge. These are all electron-like Fermi
surfaces. Figures.~\ref{smsc}c and \ref{smsc}d show the photoemission intensity along
$\Gamma$-M for the superconducting and semiconducting phases
respectively. For the superconducting phase, an indirect gap of about
25~meV could be observed between the band top of $\alpha$ and $\beta$ at $\Gamma$ and the
bottom of the $\delta$ band at M (Fig.~\ref{smsc}c). Coincidentally, the chemical potential
of the semiconductor happens to be situated in this gap. As shown in Fig.~\ref{smsc}d, the band top of $\alpha$ and $\beta$ shifts up by 55~meV compared with that in the superconductor, and is thus 20~meV below $E_F$ at $\Gamma$ in the semiconducting sample (Fig.~\ref{smsc}d). The $k_z$ dependence of the electronic structure is remarkable in the superconducting phase, but is negligible in the semiconducting phase. Along the Z-A direction, although the indirect gap below $E_F$ could not be clearly observed in Fig.~\ref{smsc}e for the superconducting phase, again there is a gap of 20~meV between the $E_F$ and the band top of $\alpha$ and $\beta$ for the semiconducting phase(Fig.~\ref{smsc}f). Note that, the $\delta$ and $\kappa$ bands are absent near $E_F$ in semiconducting samples. One could have deduced the unoccupied $\delta$ bands to be the dashed curves in Fig.~\ref{smsc}b, if it had shifted the same amount as the $\beta$ band at M. However, no traces of bands above $E_F$ in the semiconducting sample were detectable up to 200~K, which means that the unoccupied states are too far to be populated by the temperature broadening of the Fermi-Dirac distribution. Therefore, we could safely deduce a lower limit of the band gap of 40~meV in the semiconducting sample (Fig.~\ref{smsc}b), which is consistent with the small indirect gap of about 30~meV deduced from the optical conductivity data of the semiconducting K$_{0.8}$Fe$_{2-y}$Se$_2$ \cite{WangNLKFe2Se2}.

The evolution of the superconducting and semiconducting phases is summarized in Fig.~\ref{smsc}b. Our results strongly suggest that the \textbf{actual} parent compound of K$_{x}$Fe$_{2-y}$Se$_2$ superconductor is not the insulating phases, but rather the semiconducting phase. With electron doping, the semiconductor would evolve into a superconductor. In contrast to the Mott-insulating state in cuprates and the metallic SDW state in other iron-pnictide high-$T_c$'s, the semiconducting phase of K$_{x}$Fe$_{2-y}$Se$_2$ is characterized by the fully occupied band structure with a small band gap and no magnetic order as discussed below, which suggests that the U does not play a dominating role here. However, there is non-trivial correlation effect involved in this semiconducting phase. For example, the semiconductor has much weaker $k_z$ dependence compared with the superconductor, and the doping affects the  $\omega$ band in a much more pronounced way than the other bands. Such an interesting semiconducting phase was largely neglected in previous theoretical and experimental studies. Our findings identify this semiconducting phase as a novel starting point to model the superconductivity in K$_{x}$Fe$_{2-y}$Se$_2$, which is rather unique compared with other high-$T_c$'s.

We note that the earlier TEM studies have found that both vacancy ordered
phase and disordered phase exist in different regions of the sample
\cite{TEM1}. It was speculated that the vacancy-ordered phases is insulating, while the vacancy-disordered phase is superconducting. Consistently, sharp nuclear magnetic resonance (NMR) observed in K$_x$Fe$_{2-y}$Se$_2$ suggests that the superconducting phase does not coexist with magnetic order\cite{Imai}. On the other hand,  neutron scattering, M$\ddot{o}$ssbauer spectroscopy, muon spin rotation and relaxation, and transport measurements strongly suggest that the superconductivity coexists with the antiferromagnetic order, where each four-iron spin block even has a total moment as large as 13$\mu_B$ \cite{Bao1, CanM, coex2, coex3}. However surprisingly, recent TEM study suggests that the vacancy order disappears when entering the superconducting state at low temperatures \cite{TEM2}, which is hard to reconcile with the large energy scale involved for the vacancy order, but could provide a natural explanation to the NMR data. To date, there are still controversies and unsolved puzzles on whether the magnetic order and the superconductivity coexist in K$_x$Fe$_{2-y}$Se$_2$.

Our detailed ARPES results provide a compelling electronic picture to resolve these issues. We do not observe any folded feature related to the strong vacancy or magnetic order in the low energy electronic structure in Fig.~\ref{smsc}, despite clear diffraction pattern for the $\sqrt{5} \times \sqrt{5}$ vacancy order have been observed with TEM in both the superconducting and semiconducting samples (Fig.~\ref{sep}). This striking fact together with the observed charging effects show that the superconducting/semiconducting compounds are made of two phases: the insulating AFI1 phase, where the magnetic and vacancy orders are strongly tied with each other \cite{hu}, and the superconducting/semiconducting phase without such orders. The disappearance of the vacancy order in the recent TEM study is due to charging as well. That is, the Bragg peaks corresponding to the vacancy order was smeared out by accumulated charges on the insulating domains, although the vacancy order did not disappear at low temperatures. Our findings further explain various anomalous properties of K$_x$Fe$_{2-y}$Se$_2$, such as the very low charge carrier density observed in the superconducting sample in optical studies \cite{WangNLgap}. Furthermore, it may also hint that the hump of resistivity at 120~K for the superconducting compound is just a consequence of the competition between the metallic and insulating regions in the same sample.

To summarize, we have identified four different phases of K$_x$Fe$_{2-y}$Se$_2$, and provided electronic evidence for a mesoscopic phase separation of the superconducting/semiconducting phase and one of the insulating phases. We illustrated a Mott-like spectral weight transfer in AFI1 and AFI2 phases, which shows that U becomes evident when the magnetic order sets in. On the other hand, except a strong renormalization of bandwidth, the electronic structures measured in superconducting/semiconducting phases are not drastically altered from LDA results. The semiconducting phase is most likely the actual parent compound that will be led to the superconductor upon electron doping. Our results not only give a comprehensive understanding of various anomalous properties of this material, but also provide the foundation for a microscopic understanding of this new series of iron-based superconductors.

\textbf{Methods}

K$_x$Fe$_{2-y}$Se$_2$ single crystals were synthesized by self-flux method as described elsewhere in detail \cite{ChenPhase}, which show flat shiny surfaces with dark black color. The superconducting sample shows the onset superconducting transition temperature ($T_c$) of 31.7~K, and it reaches zero resistivity at 31.2~K.   The actual chemical compositions of the samples under study were determined by energy dispersive X-ray (EDX) spectroscopy, which gives K$_{0.77}$Fe$_{1.65}$Se$_2$ for the superconductor, K$_{0.65}$Fe$_{1.67}$Se$_2$ for the semiconductor,  K$_{0.78}$Fe$_{1.59}$Se$_2$ for the antiferromagnetic insulator phase I (AFI1), and K$_{0.95}$Fe$_{1.61}$Se$_2$ for the antiferromagnetic insulator phase II (AFI2) respectively. The TEM diffraction experiment was performed with JEOL JEM-2100F  at room temperature. The in-house ARPES measurements were performed with  21.2~eV He-I$\alpha$  light from a SPECS UVLS discharge lamp. The synchrotron ARPES experiments were performed at Beamline 5-4 of SSRL synchrotron facility.  All the data were taken with Scienta electron analyzers, the overall energy resolution is 15~meV in-house or 10~meV at SSRL, and angular resolution is 0.3 degree. The samples were cleaved \textit{in situ}, and measured under ultra-high-vacuum of $5\times10^{-11}$\textit{torr}.

\textbf{Author contributions}

F.C., M.X., and Q.Q.G. contribute equally to this work. M.X., Q.Q.G., F.C., Z.R.Y., Y.Z., L.X.Y., and B. P. X. performed ARPES measurements. J.J. conducted sample characterizations. F.C., J.J., M.Z., A.F.W. and X.H.C. provided the samples. R.C.C. conducted TEM measurements. M.X., Q.Q.G., Y.Z., D.W.S., D.L.F. and J.P.H. analyzed the ARPES data. D.L.F., X.M.X., and M.H.J. are responsible for the infrastructure. D.L.F. and J.P.H. wrote the paper. D.L.F. is responsible for project direction and planning.

\textbf{Acknowledgement: }We are grateful to Dr. Donghui Lu for experimental assistance at SSRL, and Prof. Tao Xiang, Prof. Nanlin Wang and Prof. Jianhui Dai for helpful discussions. This work is supported in part by the National Science Foundation of China, Ministry of Education of China, Science and Technology Committee of Shanghai Municipal, and National Basic Research Program of China (973 Program)  under the grant Nos. 2011CB921802 and 2011CBA00112. SSRL is operated by the US DOE, Office of Basic Energy Science, Divisions of Chemical Sciences and Material Sciences.

\end{document}